\documentclass[12pt, a4paper]{article}
\usepackage{abstract}
\usepackage{amsfonts}
\usepackage{amsmath}
\usepackage{amssymb}
\usepackage{authblk}
\usepackage[british]{babel}
\usepackage[backend=biber,bibencoding=utf8,sorting=none]{biblatex}
\usepackage{float}
\usepackage{hyperref}
\usepackage[utf8]{luainputenc}
\usepackage{csquotes} 
\usepackage{graphicx}
\usepackage{mathrsfs}
\usepackage{siunitx}
\usepackage{slashed}
\usepackage{subcaption}
\usepackage{tabularx}
\usepackage[compat=1.1.0]{tikz-feynman}
\usepackage[capitalise]{cleveref} 

\newcommand{\amplitude}{\mathcal{M}}

\newcommand{\rcite}[1]{Ref.~\cite{#1}}
\newcommand{\rscite}[1]{Refs.~\cite{#1}}

\newcommand{\D}[1]{{\Delta(#1)}}
\newcommand{\diracSigma}{\sigma}
\newcommand{\dsigmadm}{\mathrm{d}\sigma/\mathrm{d}\invM{}}

\newcommand{\eminus}{\mathrm{e}^{-}}
\newcommand{\eplus}{\mathrm{e}^{+}}

\newcommand{\gev}[1]{\SI{#1}{\giga\electronvolt}}

\newcommand{\hadronTensor}{W}
\newcommand{\hermitianConjugate}{\mathrm{h.c.}}

\renewcommand{\Im}{\operatorname{Im}}
\newcommand{\im}{i}
\newcommand{\invM}{m_\mathrm{inv}}
\newcommand{\iu}{i}

\newcommand{\lagrangian}{\mathscr{L}}
\newcommand{\leptonTensor}{l}
\newcommand{\li}[1]{{#1}}

\newcommand{\mt}{g}
\newcommand{\mub}[1]{\SI{#1}{\mu\mathrm{b}}}

\newcommand{\neutron}{\mathrm{n}}
\newcommand{\N}[1]{{\nucleon(#1)}}
\newcommand{\nucleon}{\mathrm{N}}

\newcommand{\photon}{\gamma}
\newcommand{\photonNR}{\photon\nucleon\resonance}
\newcommand{\piminus}{\pi^-}
\newcommand{\piNR}{\pion\nucleon\resonance}
\newcommand{\pion}{\pi}
\newcommand{\piplus}{\pi^+}
\newcommand{\pivector}{\vec{\pi}}
\newcommand{\proton}{\mathrm{p}}
\newcommand{\psibar}{\overline{\psi}}
\newcommand{\Psibar}{\overline{\Psi}}

\renewcommand{\Re}{\operatorname{Re}}
\newcommand{\reaction}{\pi^-\proton\to\neutron\eplus\eminus}
\newcommand{\reactionpipi}{\pi^-\proton\to\neutron\piplus\piminus}
\newcommand{\resonance}{\mathrm{R}}
\newcommand{\dalitzdecay}{\resonance\to\nucleon\eplus\eminus}
\newcommand{\rhodecay}{\rho^\mathrm{lep}}
\newcommand{\rhoMeson}{\rho}
\newcommand{\rhoMesonText}{$\rhoMeson$ meson}
\newcommand{\rhoMesonDash}{$\rhoMeson$-meson}
\newcommand{\rhoMesonsText}{$\rhoMeson$ mesons}
\newcommand{\rhoNR}{\rhoMeson\nucleon\resonance}
\newcommand{\rhovector}{\xvec{\rhoMeson}}
\newcommand{\rhoprod}{\rho^\mathrm{had}}
\newcommand{\rhophase}{\phi_{\rhoNR}}

\newcommand{\tauvector}{\vec{\tau}}
\newcommand{\Thetavector}{\vec{\Theta}}

\newcommand{\varGammaFive}{\Gamma}
\newcommand{\varGammaFiveTilde}{\tilde{\Gamma}}

\newcommand{\oneHalf}{{1/2}}
\newcommand{\threeHalves}{{3/2}}

\newcommand{\couplingConstant}{g}
\newcommand{\couplingConstantPiNR}{\couplingConstant_{\pion\nucleon\resonance}}
\newcommand{\couplingConstantRhoNR}{\couplingConstant_{\rhoMeson\nucleon\resonance}}
\newcommand{\couplingConstantGammaNR}{\couplingConstant_{\photon\nucleon\resonance}}

\makeatletter
\newlength\xvec@height%
\newlength\xvec@depth%
\newlength\xvec@width%
\newcommand{\xvec}[2][]{%
	\ifmmode%
	\settoheight{\xvec@height}{$#2$}%
	\settodepth{\xvec@depth}{$#2$}%
	\settowidth{\xvec@width}{$#2$}%
	\else%
	\settoheight{\xvec@height}{#2}%
	\settodepth{\xvec@depth}{#2}%
	\settowidth{\xvec@width}{#2}%
	\fi%
	\def\xvec@arg{#1}%
	\def\xvec@dd{:}%
	\def\xvec@d{.}%
	\raisebox{.2ex}{\raisebox{\xvec@height}{\rlap{%
				\kern.05em
				\begin{tikzpicture}[scale=1]
				\pgfsetroundcap
				\draw (.05em,0)--(\xvec@width-.05em,0);
				\draw (\xvec@width-.05em,0)--(\xvec@width-.15em, .075em);
				\draw (\xvec@width-.05em,0)--(\xvec@width-.15em,-.075em);
				\ifx\xvec@arg\xvec@d%
				\fill(\xvec@width*.45,.5ex) circle (.5pt);%
				\else\ifx\xvec@arg\xvec@dd%
				\fill(\xvec@width*.30,.5ex) circle (.5pt);%
				\fill(\xvec@width*.65,.5ex) circle (.5pt);%
				\fi\fi%
				\end{tikzpicture}%
			}}}%
			#2%
		}
		\makeatother

		\renewcommand{\vec}[1]{\xvec[]{#1}}

\title{Role of baryon resonances in the $\reaction$ reaction within an effective-Lagrangian model}

\author[1]{Miklós Zétényi}
\author[2]{Deniz Nitt}
\author[2]{Michael Buballa}
\author[2,3]{Tetyana Galatyuk}
\affil[1]{Wigner Research Centre for Physics, Budapest, Hungary}
\affil[2]{Technische Universität Darmstadt, Department of Physics, Darmstadt, Germany}
\affil[3]{GSI Helmholtzzentrum für Schwerionenforschung GmbH, Darmstadt, Germany}
\begin{document}

\maketitle
\begin{abstract}
We present a study of the reaction $\reaction$ for $\sqrt{s}=\SI{1.49}{\giga\electronvolt}$, including non-resonant Born terms and contributions of the $\N{1440}$, $\N{1520}$, $\N{1535}$ resonances ($\resonance$), using an effective-Lagrangian model, which we extended by a phenomenological phase factor at the $\resonance\nucleon\rhoMeson$ vertex function. We give predictions for both the differential cross section $\dsigmadm$ and the spin density matrix elements of the virtual photon that decays into the lepton pair. In the studied energy range, the cross section is dominated by the Born and $\N{1520}$ contributions.
\newline
\newline
\end{abstract}

\section{Introduction}
Electromagnetic probes (photons and correlated lepton-antilepton pairs, called dileptons) provide a unique tool to study properties of strongly interacting matter \cite{Rapp:2009yu,Gale:2009gc,Rapp:2014hha}.
Once produced in pion-, proton- or nucleus-nucleus collisions, photons and dileptons leave the interaction volume essentially unaffected by final-state interactions. 
Thus they preserve information about the medium they were created in, including the matter present in the hot and/or dense stage of the reaction \cite{Adamczewski-Musch:2019byl,Arnaldi:2006jq,Adamova:2006nu,Makek:2016rnn,Acharya:2018nxm}.
In the low-mass region ($\invM < \gev{1}$), excess dileptons are radiated from channels involving baryons \cite{Herrmann:1993za,Chanfray:1993ue,Peters:1997va,Rapp:1995zy,Cabrera:2000dx}.
A strong broadening of the $\rho$ meson has been observed in experiments spanning from a few GeV to few TeV energy range.
This trend hints at a strong coupling of the $\rho$ meson to baryonic resonances, being manifestations of in-medium modifications of vector mesons \cite{Rapp:2011is}.
A determination of the couplings of baryonic resonances to final states involving dileptons is particularly important for the understanding of the dilepton emissivity of hot and dense nuclear matter.

Dileptons produced in hadronic processes test the electromagnetic interaction of hadrons in a kinematical domain inaccessible in electro- or photoproduction experiments. In pion-nucleon collisions,  an important contribution is given by processes where a baryon resonance is excited and subsequently decays into the final-state particles.
The resulting dilepton spectrum gives access to electromagnetic transition form factors of these baryon resonances in the timelike region. These form factors have been studied for the case of $\D{1232}$, $\N{1520}$ and $\N{1535}$ in the framework of the covariant spectator quark model which takes into account contributions from the valence quark core and the meson cloud \cite{Ramalho:2015qna,Ramalho:2016zgc,Ramalho:2020nwk}.

The angular distribution of the lepton pair reflects the polarization state of the virtual photon, which in turn is determined by the hadronic reaction in which it was created \cite{Speranza_2017}.
Neutral vector mesons have the same quantum numbers as photons, therefore they can convert to a virtual photon. According to the vector-meson dominance (VMD) hypothesis \cite{Sakurai:1960ju}, an essential contribution to the electromagnetic interaction of hadrons proceeds through an intermediate neutral vector meson, whose properties can be studied in the dilepton invariant-mass spectrum.

The HADES collaboration has recently carried out experiments using secondary pions impinging on polyethylene and carbon targets. The beam was provided by the SIS18 heavy-ion synchrotron at GSI in Darmstadt, Germany.
The results are extracted for the $\piminus\proton$ reaction at the center-of-mass (c.m.) energy of $\sqrt{s}=\gev{1.49}$, which lies in the second resonance region. HADES results for pion-pair production in $\piminus\proton$ collisions are presented in \rcite{HADESpipi}. The \rhoMesonText{}, which predominantly decays to a pion pair, contributes strongly to the reaction $\reactionpipi$. Similarly to dileptons, the angular distribution of the pion pair reflects the polarization state of the decaying \rhoMesonText{}.

In this paper we present a model calculation of the dilepton production process in pion-nucleon collisions at the c.m.\ energy of $\sqrt{s}=\gev{1.49}$. We give predictions for the invariant-mass spectrum of dielectrons. We also point out how the spin density matrix of the virtual photon 
could be reconstructed based on the angular distribution of lepton pairs. Our studies are based on the previous models used in Refs.~\cite{Speranza_2017} and \cite{mikloswolf}.  

In \rcite{mikloswolf} the reaction $\reaction$ was studied in an effective-Lagrangian model taking into account both the contributions with baryon resonances in the $s$- and $u$-channel, and the non-resonant (Born) terms. Form factors were included for both the resonance and the Born contributions. A version of the VMD model was employed where electromagnetic interactions of hadrons can proceed both via an intermediate \rhoMesonText{} and a direct coupling to the photon field. Predictions for the dilepton invariant-mass spectrum were presented for various beam energies.

In \rcite{Speranza_2017} the focus was on the angular distribution of dileptons, which was studied in terms of anisotropy coefficients. The main conclusion was that the angular distribution of dileptons is characteristic of the creation process, in particular the spin-parity of intermediate baryon resonances. In that study, Born contributions were not considered, while interactions of higher-spin baryons were treated according to the consistent interaction scheme described in \rcite{Vrancx_2011}.

In the present calculation, we describe baryon resonances in the same way as in \rcite{Speranza_2017}. We include Born contributions and employ form factors as described in \rcite{mikloswolf}. We extended the models by including a phenomenological phase factor in the Lagrangians describing the resonance transition to a nucleon and \rhoMesonText{}. This factor influences the numerical predictions of the model via interference effects. 

More details about our model will be given in \cref{section:model} and in the Appendix. Our results will be presented in \cref{section:results}. Finally, in \cref{section:conclusions}, we draw our conclusions and discuss possible future directions. 

\section{Model}
\label{section:model}
In this Section, we briefly review the main elements of the applied models and outline the calculation of the observables. For further details the interested reader is referred to \rscite{Speranza_2017} and \cite{mikloswolf}.

\subsection{Kinematics and Observables}
The differential cross section of the process $\reaction$ is given by
\begin{equation}
  \label{eq:dsdm}
  \frac{\mathrm{d}\sigma}{\mathrm{d}\invM \mathrm{dcos}\theta_{\gamma^*} \mathrm{d}\Omega_{e}} = 
  \frac{\invM}{64(2\pi)^4 s}\frac{|\mathbf{p}_f|}{|\mathbf{p}_i|}  
  \frac{1}{n_\text{pol}} \sum_\text{pol}
  \left\vert\mathcal{M}\right\vert^2,
\end{equation}
where $\invM$ denotes the mass of the virtual photon (= invariant mass of the $\eplus\eminus$ pair), $\theta_{\gamma^*}$ is the virtual-photon angle in the c.m.\ frame with respect to the pion-beam direction, i.e., the scattering angle, $\mathrm{d}\Omega_{e}$ is the solid angle of the final-state electron given in the virtual photon rest frame, $s$ is the square of the c.m.\ energy, $\mathbf{p}_i$ and $\mathbf{p}_f$ are the c.m.\ three-momenta of the incoming and outgoing nucleon, respectively, and $n_\text{pol} = 2$ is the number of polarization states in the incoming channel. The sum runs over the polarization states of all incoming and outgoing particles. 

Assuming that the virtual photon and final-state nucleon do not interact with each other, we can separate the hadronic and leptonic parts of the process by cutting the Feynman diagrams at the virtual photon line. 
The spin averaged squared amplitude can then be calculated as
\begin{equation}
\label{eq:squared_amplitude}
\sum \limits_{\mathrm{pol}} \left|\amplitude\right|^2 = \sum \limits_{\lambda,\lambda^\prime} \rhoprod_{\lambda\lambda^\prime}\rhodecay_{\lambda^\prime\lambda},
\end{equation}
with the spin density matrices as introduced in
\rcite{Speranza_2017},
\begin{align}
\label{eq:spin_density_matrices}
  \rhoprod_{\lambda,\lambda^\prime}(s,\invM,\theta_{\gamma^*}) &= \frac{e^2}{k^4}\epsilon_\li{\mu}(k, \lambda) \hadronTensor^\li{\mu\nu} \epsilon^*_\li{\nu}(k, \lambda^\prime), \\
  \rhodecay_{\lambda^\prime,\lambda}(\invM,\Omega_{e}) &= \epsilon_\li{\mu}(k, \lambda^\prime) \leptonTensor^\li{\mu\nu} \epsilon^*_\li{\nu}(k, \lambda),
\end{align}
describing the two sub-processes.
The tensors $\hadronTensor^\li{\mu\nu} = \sum \limits_\text{pol} \amplitude_\text{had}^\li{\mu}\amplitude_\text{had}^\li{\nu*}$ and $\leptonTensor^\li{\mu\nu} = \sum \limits_\text{pol} \amplitude_\text{lep}^\li{\mu}\amplitude_\text{lep}^\li{\nu*}$ are related to the hadronic and the leptonic part of the amplitude, respectively, with 
$\amplitude_\text{had}^\li{\mu}$ and
$\amplitude_\text{lep}^\li{\mu}$ denoting the 
corresponding transition currents.
The virtual photon is characterized by its four-momentum $k$ and polarization $\lambda$, which also enter the polarization vector $\epsilon_\li{\mu}(k, \lambda)$. 

As pointed out in \rcite{Speranza_2017}, the form of $\rhodecay_{\lambda^\prime,\lambda}$ can be calculated from quantum electrodynamics, therefore the angular distribution of $\eplus\eminus$ pairs can be expressed in terms of the hadronic density matrix elements.
Making use of the hermiticity of $\rhoprod_{\lambda,\lambda^\prime}$ we obtain for the squared amplitude
\begin{align}
    \label{eq:ang_dist}
    \sum \limits_{\mathrm{pol}} \left|\amplitude\right|^2 &\propto
    (1+\cos^2\theta_{e})\left(\rhoprod_{1,1} + \rhoprod_{-1,-1}\right)
    + 2\sin^2\theta_{e}\rhoprod_{0,0} \nonumber \\
    &+ \sqrt{2}\sin 2\theta_{e}
    \left[\cos\phi_{e}(\Re\rhoprod_{1,0}-\Re\rhoprod_{-1,0})\right. \nonumber \\
    &+ \left. \sin\phi_{e}(\Im\rhoprod_{1,0}+\Im\rhoprod_{-1,0}) \right] \nonumber \\
    &+ 2\sin^2\theta_{e}\left(\cos 2\phi_{e}\Re\rhoprod_{1,-1}
    +\sin 2\phi_{e}\Im\rhoprod_{1,-1}\right),
\end{align}
see Eq.\ (15) in \rcite{Speranza_2017}. This makes it possible, at least in principle, to determine the hadronic density-matrix elements $\rhoprod_{\lambda,\lambda^\prime}$ based on the angular distribution of the $\eplus\eminus$ pairs obtained in the experiments.


\subsection{Vector-Meson Dominance and Effective Lagrangian}
\label{section:VMD}
The vector-meson dominance model (VMD) was initially proposed by Sakurai to describe the coupling of a photon to a hadronic current via an intermediate vector meson \cite{Sakurai:1960ju,12853}. We employ the later version of the VMD as discussed in \rcite{Kroll:1967it}, which also allows for a direct coupling of hadrons to the electromagnetic field. The Lagrangian of this version of VMD can be symbolically written as 
\begin{equation}
    \lagrangian_{\text{VMD}} = -\frac{e}{2g_\rho} F^{\mu\nu}\rho^0_{\mu\nu} + \sum \limits_{v,w} \left( \lagrangian_{\gamma vw} + \lagrangian_{\rho vw}\right),
\label{eq:LVMD}    
\end{equation}
where $F_{\mu\nu} = \partial_{\mu}A_{\nu}-\partial_{\nu}A_{\mu}$ and $\rho^0_{\mu\nu} = \partial_{\mu}\rho^0_{\nu}-\partial_{\nu}\rho^0_{\mu}$ are the field-strength tensors of the  photon and neutral \rhoMesonText{}, respectively, expressed in terms of the corresponding fields. (In $\rho^0_{\mu\nu}$ we omitted the term quadratic in the $\rho$ field because it does not contribute to the process we study.)

The first term in Eq.~(\ref{eq:LVMD}) describes the $\rhoMeson$-$\photon$ transition and its coupling constant is controlled by the parameter  $g_\rhoMeson = 4.96$. The summation in the second term runs over pairs of hadron fields $v$, $w$, and the two symbolic terms represent both the interaction of various hadrons with the photon and the \rhoMesonText{}, like e.g.\ $\lagrangian_{\gamma\pi\pi}$ or $\lagrangian_{\rho\nucleon\nucleon}$, and transition vertices of baryon resonances ($\resonance$), like $\lagrangian_{\rhoNR}$.

According to the other version of VMD, 
\begin{equation}
    \lagrangian^{\prime}_{\text{VMD}} = -\frac{e m_{\rho}^2}{g_\rho} A^{\mu}\rho^0_{\mu} + \sum \limits_{v,w} \lagrangian^{\prime}_{\rho vw},
\label{eq:LVMDprime}    
\end{equation}
hadrons couple to the electromagnetic field only via an intermediate neutral vector meson. As pointed out in Refs.\ \cite{12853,OConnell:1995nse}, the models $\lagrangian_{\text{VMD}}$ and $\lagrangian^{\prime}_{\text{VMD}}$ are equivalent in the limit when all hadrons couple to the \rhoMesonText{} with the same universal coupling constant, which is equal to the parameter $g_\rhoMeson$ appearing in the denominator of the term describing the $\rhoMeson$-$\photon$ transition in (\ref{eq:LVMD}) and (\ref{eq:LVMDprime}), $g_{\rhoMeson\pion\pion} = g_{\rhoMeson\nucleon\nucleon} = \ldots = g_\rhoMeson$. In our model this universality does not hold, but the relative signs of the photon and \rhoMesonText{} interaction Lagrangians can be fixed based on the requirement that the two versions of VMD become equivalent when all the \rhoMesonText{} coupling constants approach a universal value.

For the non-resonant contributions we use the effective Lagrangians of \rcite{mikloswolf}. The coupling of baryon resonances to the pion, \rhoMesonDash{} and photon fields are described by the effective-Lagrangian model from \rcite{Speranza_2017}, where a consistent treatment of higher-spin contributions according to \rcite{Vrancx_2011} has been taken care of. 
An important difference between the present model and the models of \rcite{mikloswolf,Speranza_2017} is that here we fix the relative sign of hadron--photon and hadron--\rhoMesonDash{} interaction terms based on the equivalence of the two versions of VMD in the universality limit, as discussed 
above.
In \Cref{section:lagrangians} we list all terms of the effective Lagrangian applied in the present model. 
In principle, the VMD allows for all neutral vector mesons to couple between the hadronic current and the photon. Here we only consider the contribution of the $\rhoMeson^0$ meson.

For those terms in the effective Lagrangian that do not contain baryon resonances we adopt the coupling constants from \rcite{mikloswolf}.
For the resonant contributions we need to fix the coupling constants $g_{\rhoNR}$, $g_{\photonNR}$ and $g_{\piNR}$ individually.
This has been done by calculating the corresponding partial decay width of the resonance and fitting it to the experimental value. Details are specified in \Cref{sec:couplingConstants}. 
Properties of baryon resonances, including the coupling constants are given in \Cref{tbl:resonanceOverview}.


\subsection{Hadronic Processes}
\begin{figure}
  \begin{subfigure}[b]{0.16\textwidth}
    \includegraphics[width=\textwidth]{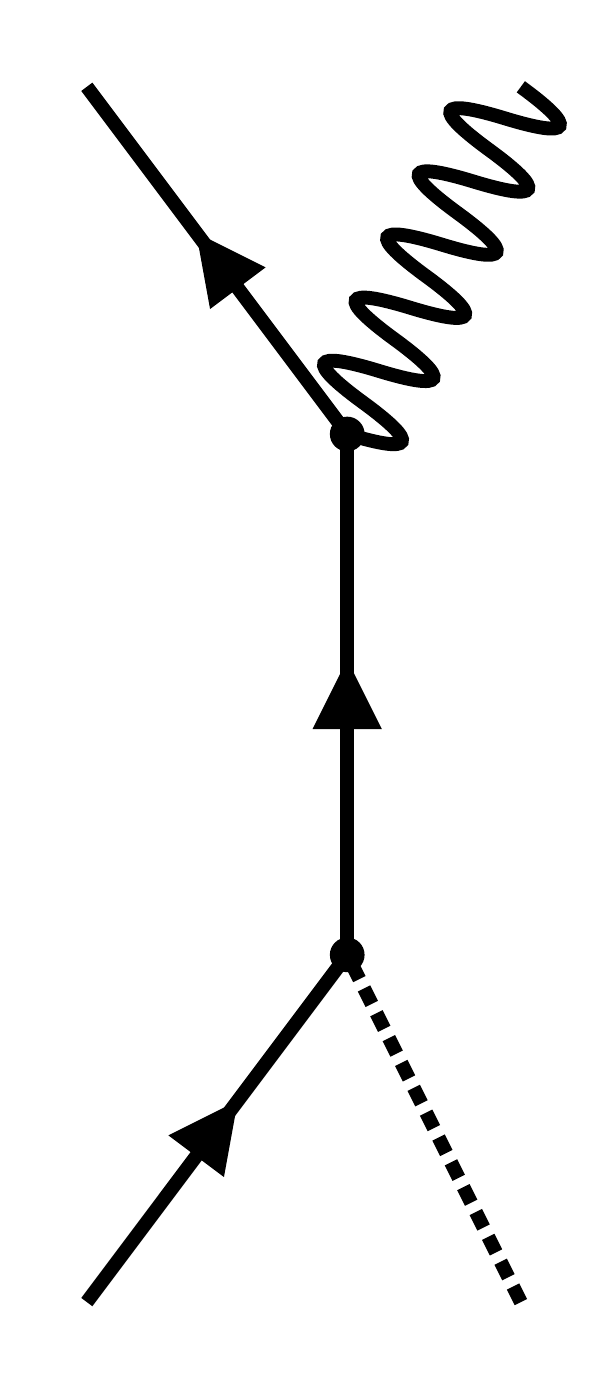}
    \subcaption{}
    \end{subfigure}
    \begin{subfigure}[b]{0.16\textwidth}
    \includegraphics[width=\textwidth]{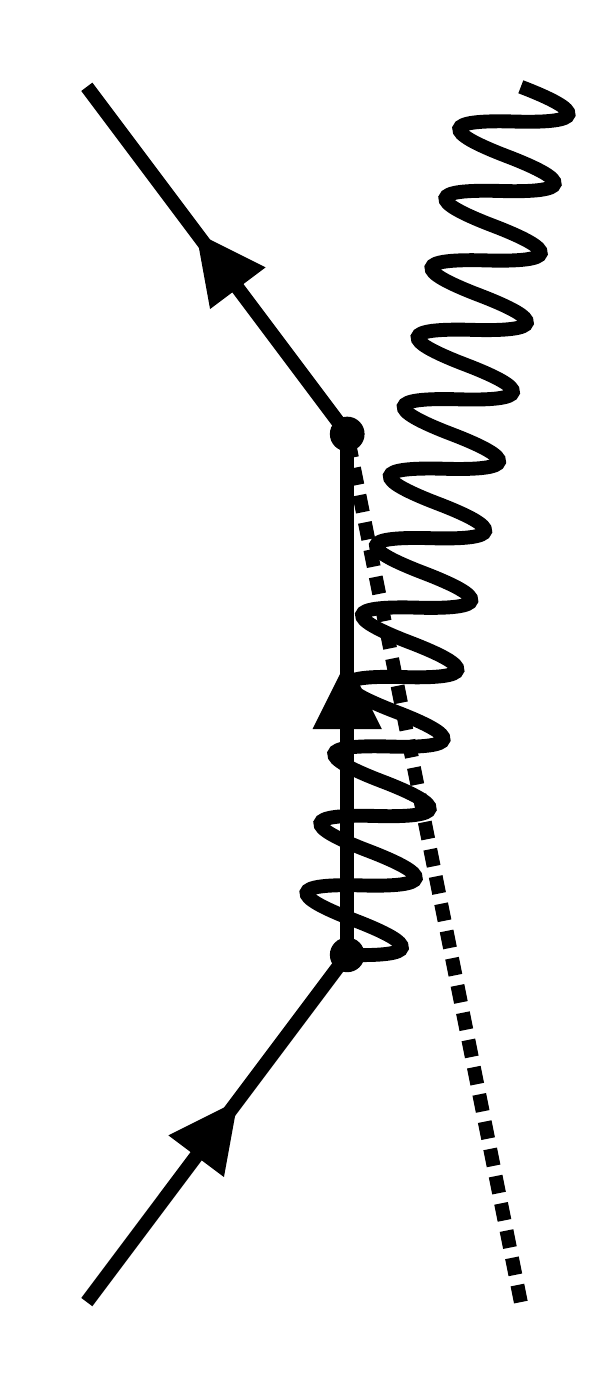}
    \subcaption{}
    \end{subfigure}
    \begin{subfigure}[b]{0.16\textwidth}
    \includegraphics[width=\textwidth]{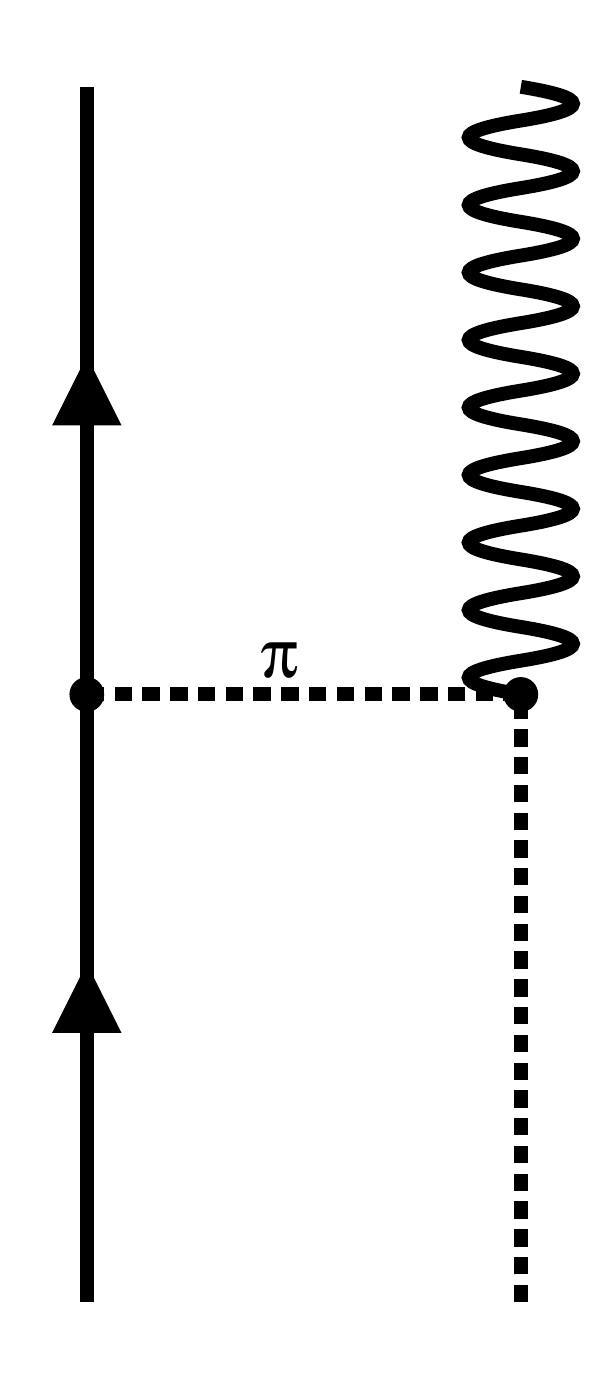}
    \subcaption{}
    \end{subfigure}
    \begin{subfigure}[b]{0.16\textwidth}
    \includegraphics[width=\textwidth]{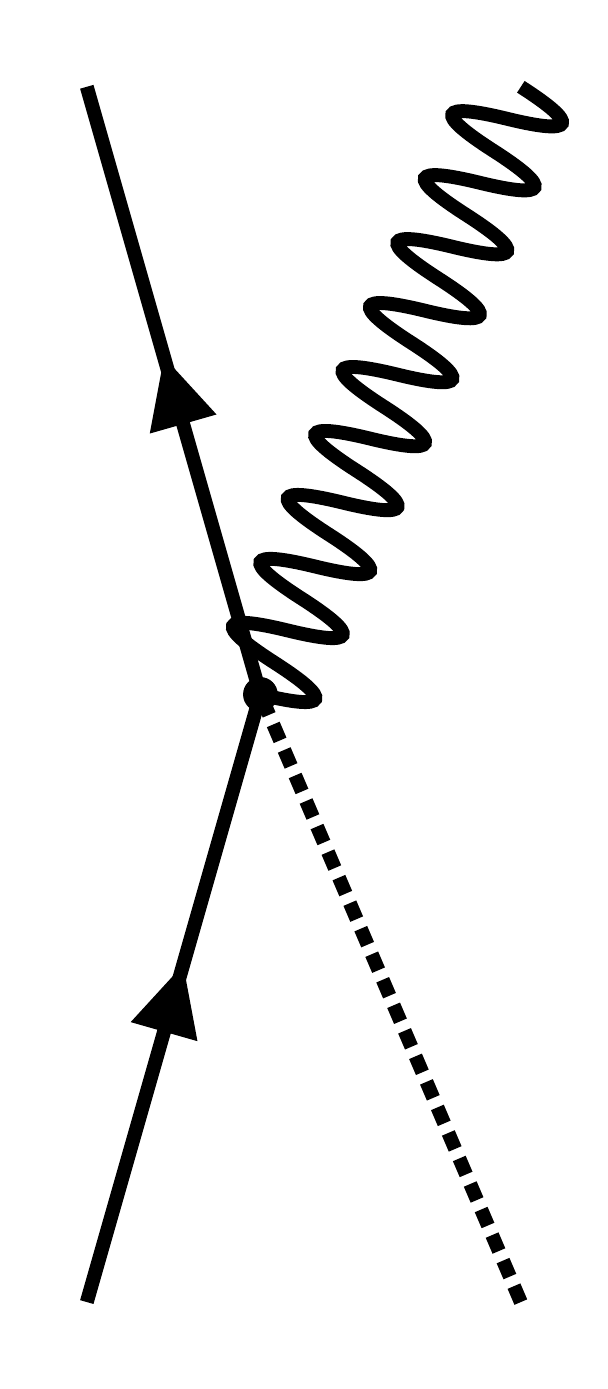}
    \subcaption{}
    \end{subfigure}
    \begin{subfigure}[b]{0.16\textwidth}
    \includegraphics[width=\textwidth]{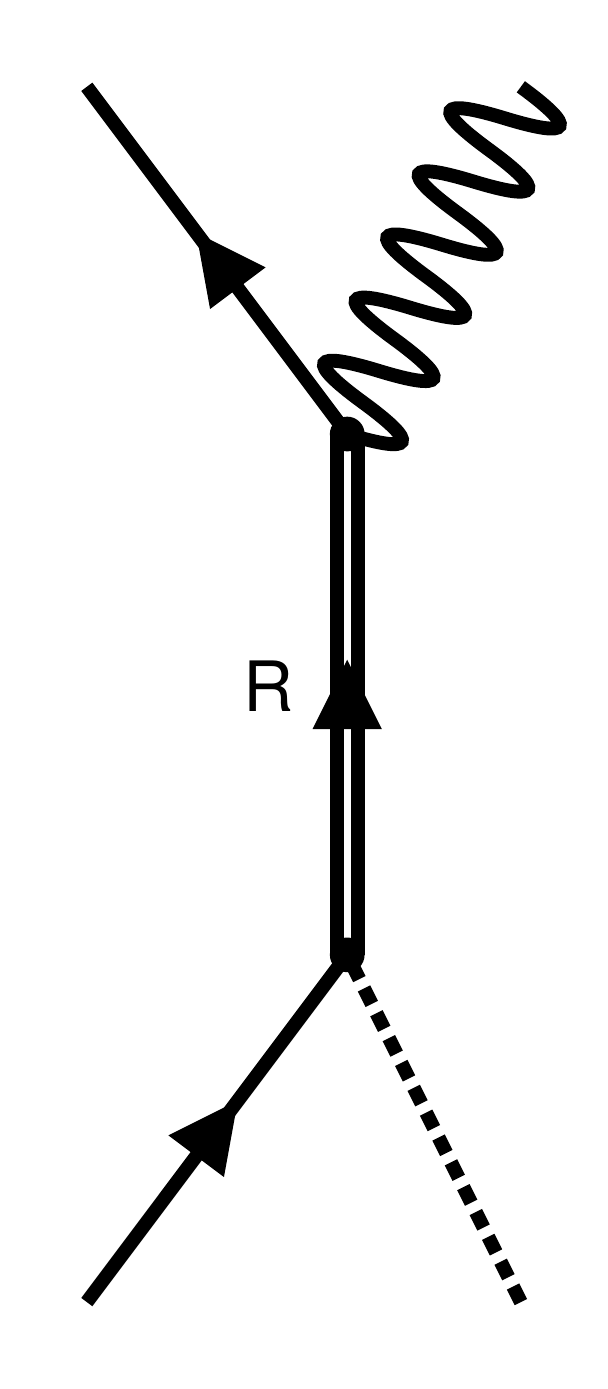}
    \subcaption{}
    \end{subfigure}
    \begin{subfigure}[b]{0.16\textwidth}
    \includegraphics[width=\textwidth]{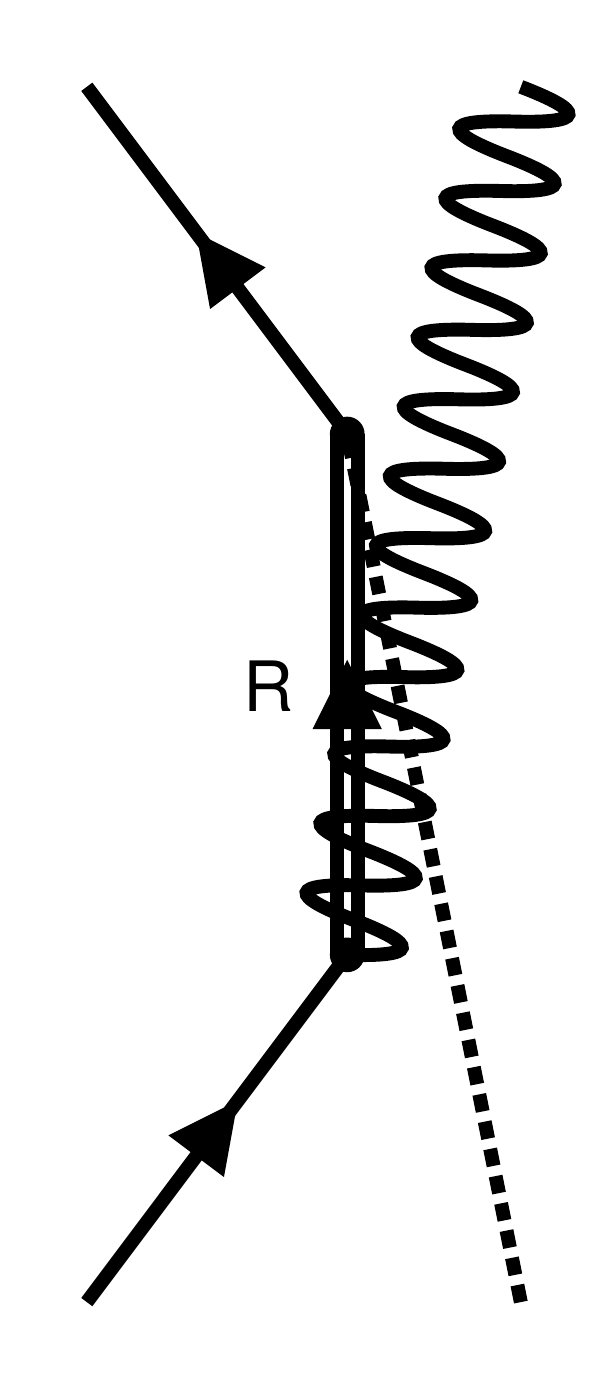}
    \subcaption{}
    \end{subfigure}
  \caption{Diagrams in lowest order contributing to the process $\reaction$. For clarity the emerging dilepton has been cut off in the Feynman diagrams. The wavy line is to be understood as the coherent sum of a direct photon and a photon coupled via an intermediate \rhoMesonText{} to the hadrons. (a)-(d) are the $s$-, $u$-, $t$-channel and contact-term Born contributions, respectively. (e) and (f) show the $s$- and $u$-channel diagrams, respectively, with an intermediate baryon resonance ($\resonance$).}
  \label{fig:feynmanDiagrams}
\end{figure}
We consider the contributions shown in \cref{fig:feynmanDiagrams} for the hadronic part of the process. These include the non-resonant $s$-, $u$- channel diagrams, the $t$-channel with an intermediate pion, together with the contact term (\cref{fig:feynmanDiagrams}~(a)-(d)) and the diagrams with $s$- and $u$-channel baryon resonances (\cref{fig:feynmanDiagrams}~(e), (f)). Each diagram in \cref{fig:feynmanDiagrams} is a sum of a direct photon and a \rhoMesonText{} contribution. 

At $\sqrt{s}=\gev{1.49}$, the nearby baryon resonances $\N{1440}$, $\N{1520}$, and $\N{1535}$ are expected to give important contributions. We also included other potential resonances in the calculations but their partial cross sections turned out to be rather small compared to the three aforementioned ones at this energy. We therefore exclude them in following discussion and the presentation of the results for the sake of clarity. 

The effective-Lagrangian model treats the complex hadron bound states as pointlike particles. The non-pointlike nature is taken into account by the inclusion of form factors. For the Born contributions we choose form factors according to the scheme described in \rcite{mikloswolf}, which preserves gauge invariance. In \rcite{Speranza_2017}, where the emphasis was on the angular distribution, no form factors were introduced. In the present study we use the form factors described in \rcite{mikloswolf} for the $\piNR$ vertices.  

The microscopic details of hadron interactions can lead to the appearance of extra phase factors at each vertex. Such phase factors have been discussed in the framework of a coupled-channel approach to meson-baryon scattering \cite{Lutz:2001mi}. In our model, we include a phase factor as a phenomenological free parameter for the vertices describing the nucleon--\rhoMesonDash{} decay mode of the baryon resonances in order to explore their effect on the interference between the direct-photon and \rhoMesonDash{} contribution of each resonant state. This essentially means that we make the corresponding coupling constants complex via the substitution $g_{\rhoNR} \to g_{\rhoNR}\exp(i\rhophase)$. This phase factor drops out when the partial decay width of resonances to the $\rhoMeson\nucleon$ final state is calculated, therefore only the moduli of the coupling constants can be determined in the way described above. The situation is different in the case of the Dalitz decay of resonances, $\dalitzdecay$, where, due to an interference between the $\rhoMeson$ and direct $\photon$ contributions, the phase factor would be relevant.

\begin{table*}
    \footnotesize
    \centering
    \begin{tabular*}{\textwidth}{c||c|c|c|c|c|c|c|c}
     & $m$ & $\Gamma$ & BR\textsuperscript{*}$\to \nucleon\pion$ & $g_{\piNR}$ & BR\textsuperscript{$\dagger$}$\to \nucleon\rho$ & $g_{\rhoNR}$ & BR\textsuperscript{*}$\to\nucleon\photon$ & $g_{\photonNR}$ \\
     & [$\si{\giga\electronvolt}$] & [$\si{\giga\electronvolt}$] & [$\%$] &  & [$\%$] &  & [$\%$] & \\ \hline
     $\N{1440}$ & $1.440$ & $0.350$ & $55-75$ & 0.38 & $< 0.2$ & 3.37 & $0.02 - 0.04$ & 0.053 \\
     $\N{1520}$ & $1.520$ & $0.110$ & $55-65$ & 0.15 & $12.2 \pm 1.9$ & 13.9 & $0.30 - 0.53$ & 0.36 \\
     $\N{1535}$ & $1.530$ & $0.150$ & $32-52$ & 0.16 & $3.2 \pm 0.7$ & 1.97 & $0.01 - 0.25$ & 0.058
    \end{tabular*}
    \caption{Resonances and their parameters presented in this paper. \textsuperscript{*} taken from \rcite{pdg} \textsuperscript{$\dagger$} taken from
    \rcite{HADESpipi}}
    \label{tbl:resonanceOverview}
\end{table*}

\section{Results}
\label{section:results}
\begin{figure*}[htb]
    \centering
    \includegraphics[width=0.8\textwidth]{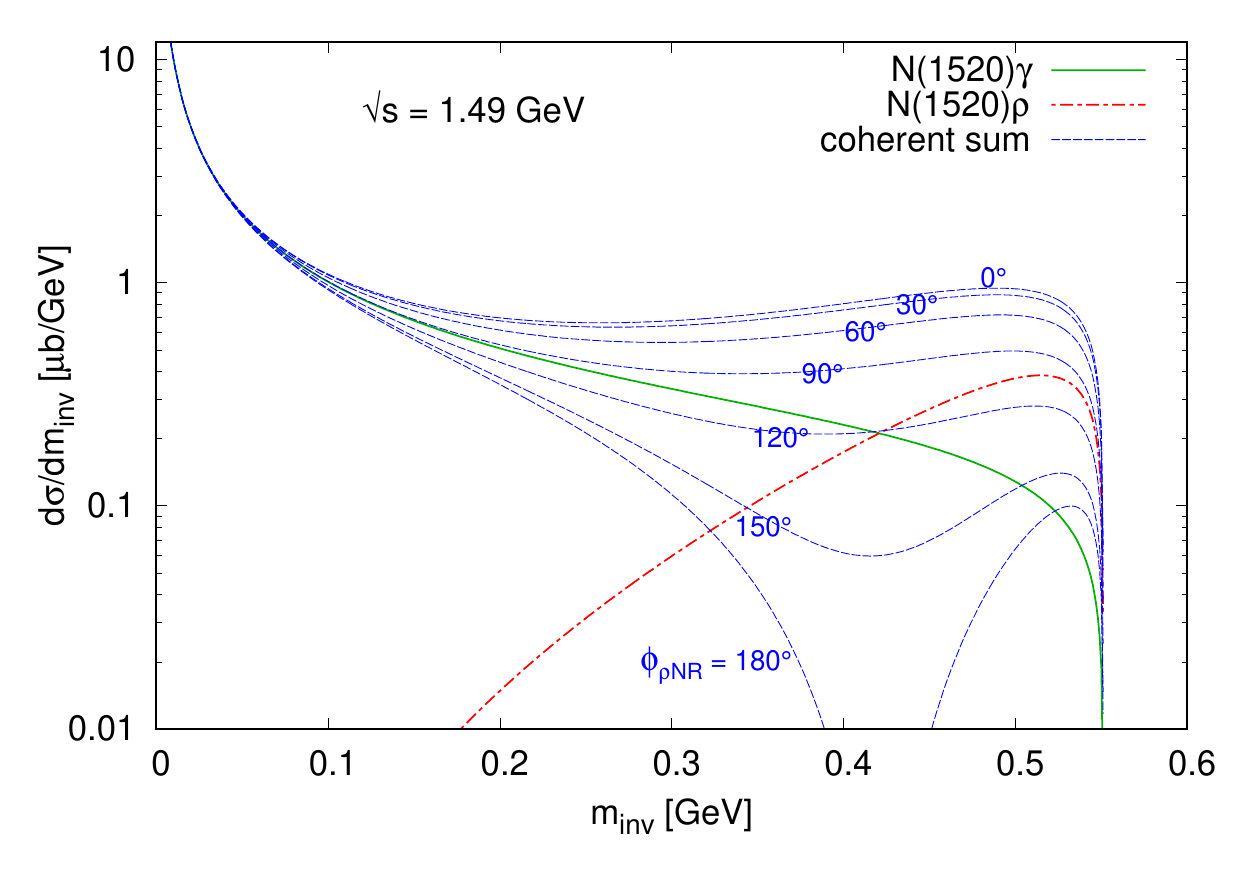}
    \caption{Contributions with an $\N{1520}$ resonance in the $s$-channel to the differential cross section $\dsigmadm$ of dilepton production. Two of the lines indicate the direct-photon and the \rhoMesonDash{} contributions. The other curves show the coherent sum of the above two, assuming different relative phases.}
    \label{fig:N1520phases}
\end{figure*}

We use the model outlined in \cref{section:model} to study the dilepton production process in pion-nucleon collisions at the energy of recent experiments by the HADES collaboration, $\sqrt{s}=\gev{1.49}$. In this energy range, the nearby $\N{1520}$ resonance is expected to strongly contribute to dilepton production due to its strong coupling to the \rhoMesonDash{}.
In \cref{fig:N1520phases} we show the $s$-channel $\N{1520}$ contribution to the differential cross section $\dsigmadm$ as a function of the dilepton invariant mass $\invM$. If we do not include any extra relative phase between the direct photon and the \rhoMesonDash{} contributions then we experience a constructive interference resulting in a smooth invariant-mass spectrum. The maximum at $\invM\approx\gev{0.5}$ is due to the \rhoMesonDash{}{} contribution. The introduction of the extra phase in the \rhoMesonDash{} contribution changes the interference pattern, and for $\rhophase=\ang{180}$ a deep minimum appears above $\invM=\gev{0.4}$.

\begin{figure*}[htbp]
    \centering
    \includegraphics[width=\textwidth]{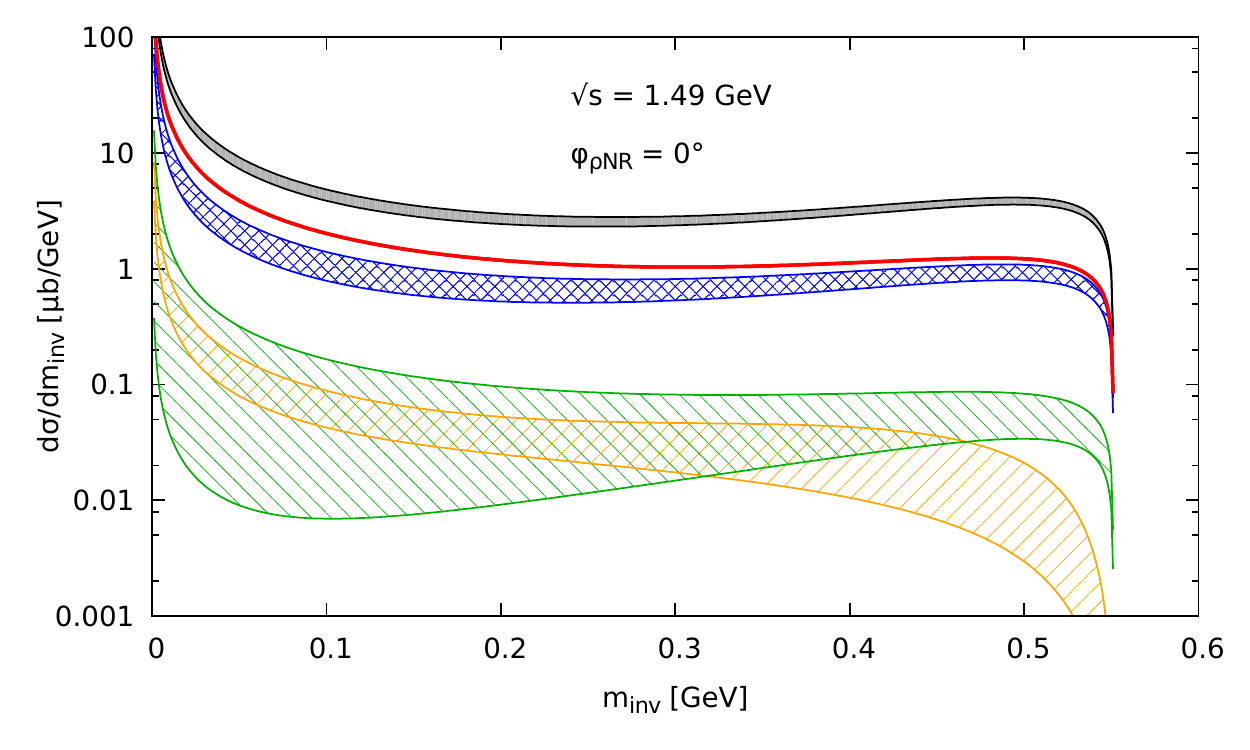}\\
    \includegraphics[width=\textwidth]{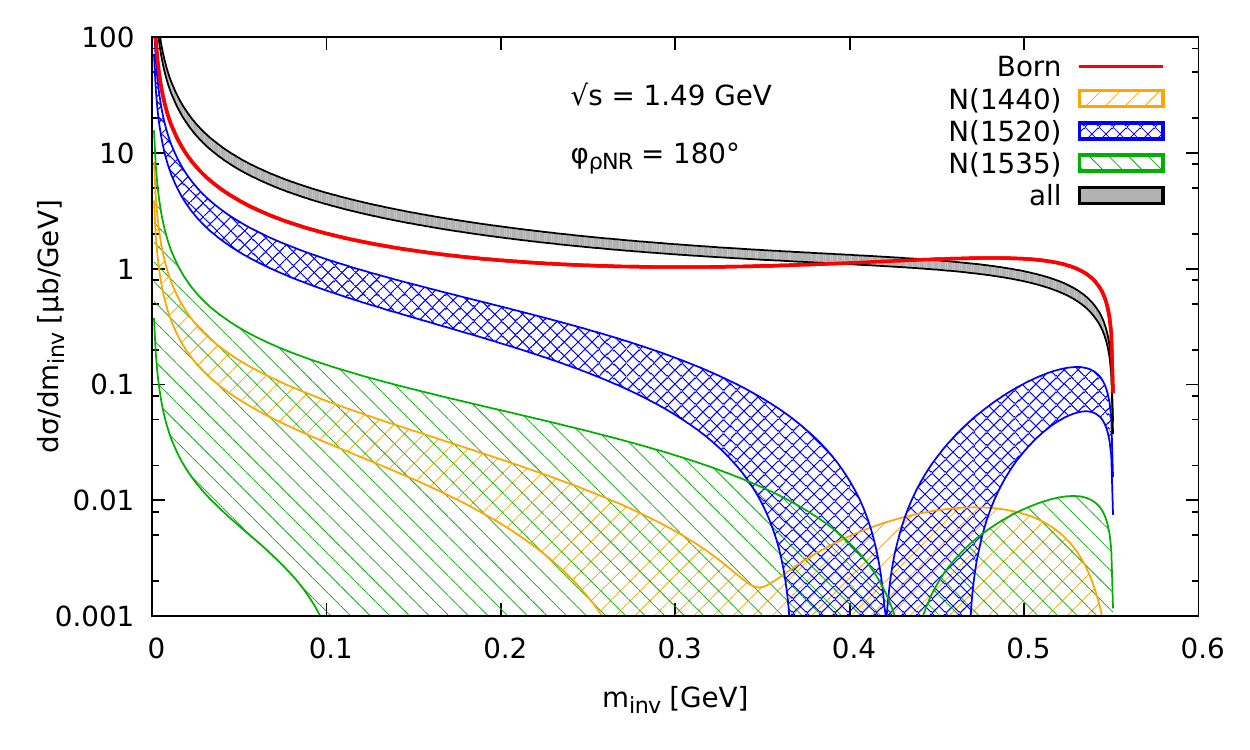}
    \caption{Differential cross section $\dsigmadm$ of the process $\pion^{-}p\to n\eplus\eminus$ at c.m.\ energy $\sqrt{s}=\gev{1.49}$, with no extra phase factor (upper plot) and with a common phase factor of $\rhophase=\ang{180}$ (lower plot) introduced in the
    $\rhoNR$ vertices. The legend for line styles and fill bands of the lower plot is valid for the upper plot too. The size of each band corresponds to the uncertainty on the width and branching ratio of the resonance as shown in \Cref{tbl:resonanceOverview}}
    \label{fig:dsigdm}
\end{figure*}

\Cref{fig:dsigdm} represents our model's prediction for the differential cross section $\dsigmadm$ of the reaction $\pion^{-}\proton\to \nucleon\eplus\eminus$ at $\sqrt{s}=\gev{1.49}$ c.m.\ energy as a function of the invariant mass $\invM$. The largest contributions are shown, i.e.\ those of the Born terms and of the baryon resonances $\N{1520}$, $\N{1535}$, and $\N{1440}$. For the resonant contributions we indicate the uncertainties arising from the errors on the resonance widths and branching ratios. Furthermore, we show the results obtained with two different assumptions on the extra phase of the resonance-$\rhoMeson$ contributions: with no extra phase factor, and with a common phase of $\rhophase=\ang{180}$ at all $\rhoNR$ vertices.

Comparing the two plots we see that the above extra phase factor 
influences both the shape and the magnitude of the differential cross section for invariant masses above \gev{0.3}. 
In the coupled-channel model of Ref.~\cite{Lutz:2001mi} relative phases of scattering amplitudes have been extracted. For instance, for the $\pion\nucleon \to \N{1520} \to \nucleon\rhoMeson$ amplitude, a phase of $\phi^{\N{1520}}_{\pion\rhoMeson} = \ang{-7.0}$ was obtained. This factor is, however, a product of the phases corresponding to the $\pion\nucleon\N{1520}$ and $\rhoMeson\nucleon\N{1520}$ vertices, while the individual phase factors cannot be extracted.
As already pointed out, we therefore treat the phase factors at the $\rhoNR$ vertices as free parameters. In principle, these phases can be different for each baryon resonance. For simplicity, we always use a common value for all resonances in this article. 
However, since by far the most important resonance contributions are due to the $\N{1520}$, relative phases for the other resonances would have only minor effects.

\Cref{fig:dsigdm} shows also the significance of the various contributions. Independent of the phase factor, the Born terms and the $\N{1520}$ provide the largest contributions. 
In the  $\rhophase = \ang{0}$ case, the $\N{1520}$ is almost as strong as the Born contribution at the large invariant mass end of the spectrum, due to its strong coupling to the \rhoMesonText{}. Integrating the cross section over the $\invM > \gev{0.3}$ region where the contribution of the \rhoMesonText{} is significant, we obtain $\sigma(\invM > \gev{0.3}) = \int_{>\gev{0.3}}(\dsigmadm) \mathrm{d}\invM  = \mub{0.73}$, which is a result of a contribution of $\mub{0.28}$ from Born terms, $\mub{0.17}$ from $\N{1520}$ terms and $\mub{0.28}$ from Born-$\N{1520}$ interference. Contributions of the other two resonances are at least an order of magnitude smaller. In the  $\rhophase = \ang{180}$ case the $\N{1520}$ contribution shows a minimum around $\invM\approx \gev{0.4}$ and becomes negligible compared to the Born term.

\begin{figure*}[htbp]
    \centering
    \includegraphics[width=0.7\textwidth]{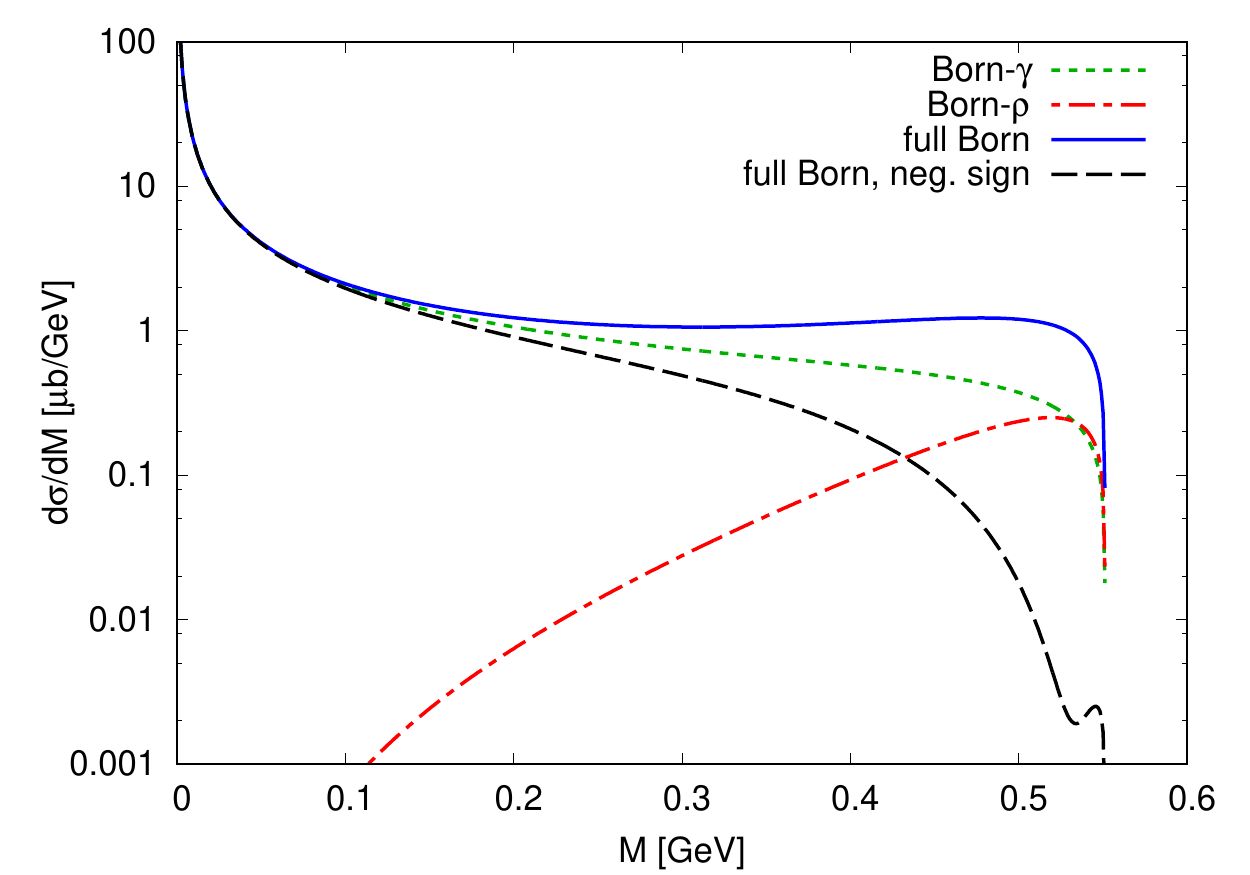}
    \caption{Born contributions to the differential cross section $\dsigmadm$ of the process $\reaction$ at c.m.\ energy $\sqrt{s}=\gev{1.49}$, within the model described in this paper (solid line) and with an extra minus sign introduced in all interaction terms involving a \rhoMesonDash (dashed line).}
    \label{fig:Born_interference}
\end{figure*}

In \rcite{mikloswolf} no attempt was made to determine the signs of hadron-$\rhoMeson$ interactions, only the signs of $\gamma\nucleon\resonance$ Lagrangians were varied in such a way that the best description of pion-photoproduction cross-sections are achieved. In fact, both for the Born terms and for the $N(1520)$ contributions, a destructive interference occurred between the photon and the \rhoMesonDash{} contributions due to the choices of signs in the relevant terms of the Lagrangian. In terms of the present model, this scenario would correspond to setting $\rhophase = \ang{180}$ and including an extra minus sign in the $\rho\nucleon\nucleon$, $\pi\rho\nucleon\nucleon$ and $\pi\pi\rho$ vertices. In Fig.~\ref{fig:Born_interference} we show the interference effects on the invariant-mass spectrum in the Born contributions. Dotted and dash-dot lines correspond to the Born-$\gamma$ and Born-$\rho$ contributions, the continuous line depicts the full Born contribution according to the model used in the present paper while the dashed line shows the full Born contribution obtained with an extra minus sign introduced in the Born-$\rho$ term. In the latter case the Born contribution is strongly reduced for large invariant masses due to the destructive interference.

\begin{figure*}[htbp]
    \centering
    \includegraphics[width=0.7\textwidth]{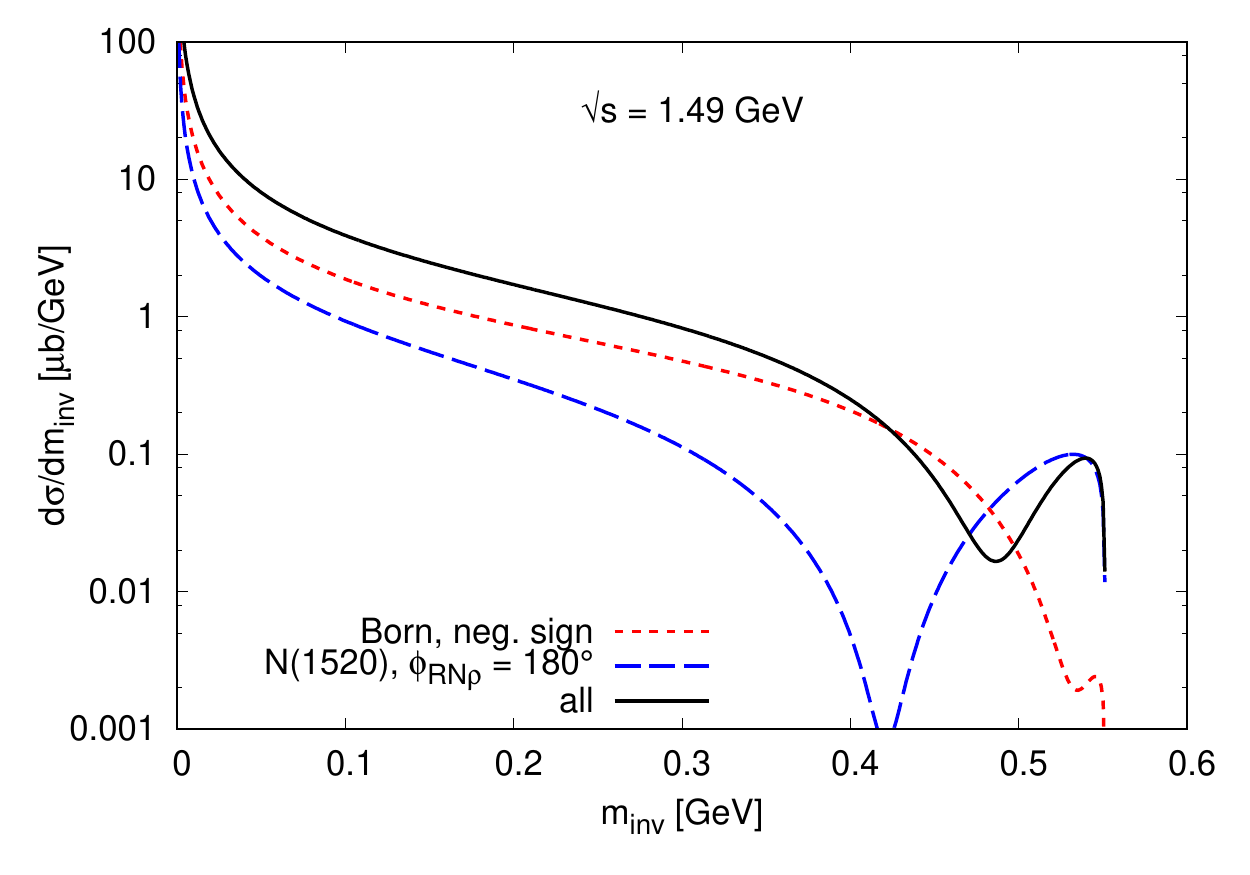}
    \caption{Differential cross section $\dsigmadm$ of the process $\reaction$ at c.m.\ energy $\sqrt{s}=\gev{1.49}$, within the model modified in such a way that destructive interference occurs between the photon and \rhoMesonDash{} contributions in the case of both the resonance and the Born terms.}
    \label{fig:dsdm_desctructive}
\end{figure*}

In Fig.~\ref{fig:dsdm_desctructive} we show the differential cross-section $\dsigmadm$ of the process $\reaction$ at c.m.\ energy $\sqrt{s}=\gev{1.49}$ as obtained from the modified model with $\rhophase = \ang{180}$ and an extra minus sign in the Born-$\rho$ contribution. In this version, the signs of interaction Lagrangians are analogous to the model of \rcite{mikloswolf}. However, the Lagrangians describing interactions of baryon resonances are different in the present model, and in the determination of coupling constants involving baryon resonances we used updated values for their masses and widths. Due to the destructive interference, the Born contribution is suppressed for high invariant mass and the $\N{1520}$ becomes dominant. As a result, the minimum in the $\N{1520}$ contribution is visible even after coherently summing all contributions. The resulting invariant mass spectrum is very similar to the one shown in \rcite{mikloswolf}.

In the following we assume no extra minus sign in the Born-$\rho$ contribution and we regard  $\rhophase = \ang{0}$ as the standard value. These choices are suggested by the arguments of \Cref{section:VMD} based on the equivalence of the two versions of VMD, and they were used in the upper plot of \Cref{fig:dsigdm}. We, however, still explore the effects of a nonzero $\rhophase$ phase factor.

In \cite{Speranza_2017}, the angular distribution of dileptons was discussed in terms of the anisotropy coefficient $\lambda_{\theta}$ and it was discussed how this and other similar coefficients are related to the polarization of the virtual photon. However, the hadronic spin density matrix, $\rhoprod_{\lambda,\lambda^\prime}$, is itself an object representing the virtual-photon polarization state. According to \cref{eq:ang_dist}, elements of this density matrix could in principle be determined based on the experimentally observed angular distribution of dileptons if sufficient statistics is available.

\begin{figure*}
    \centering
    \includegraphics[width=\textwidth]{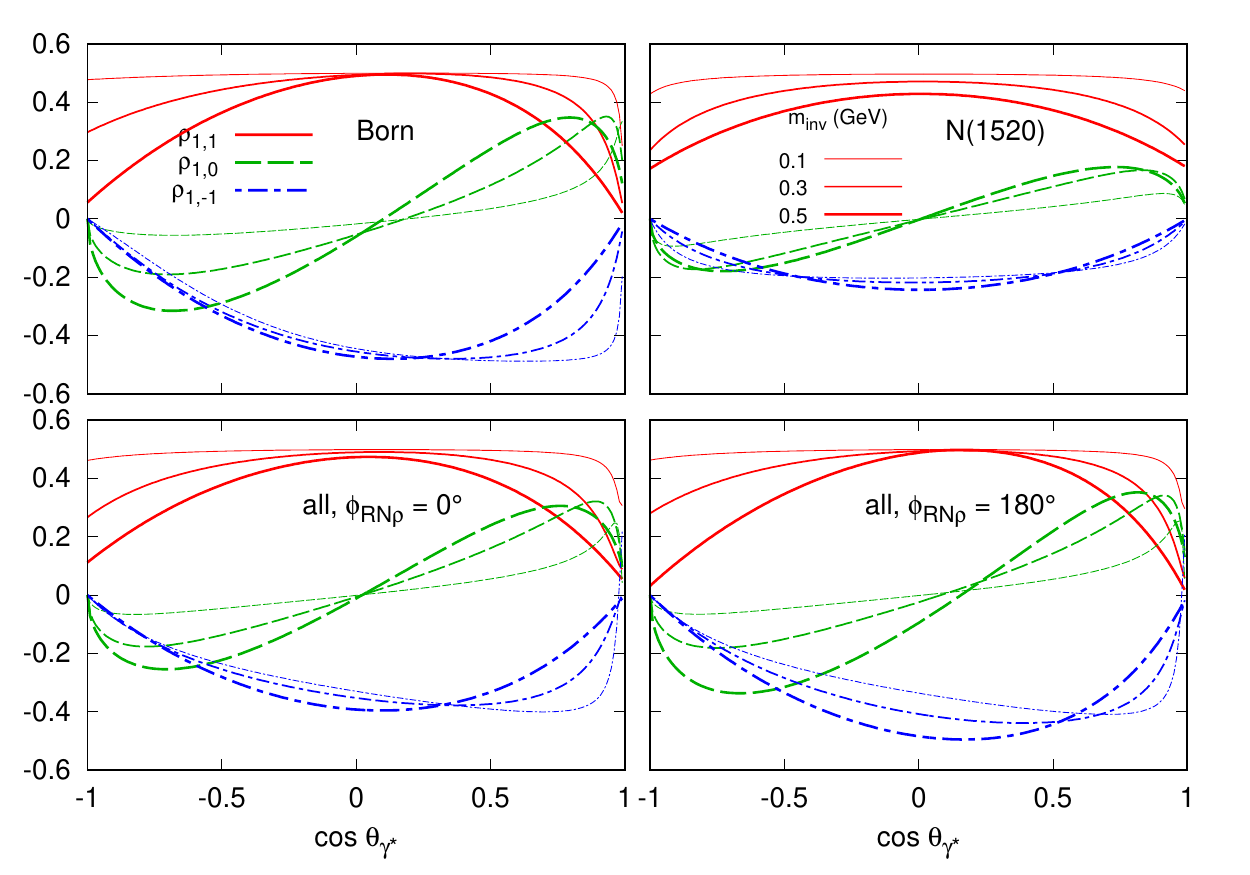}
    \caption{(color online) Hadronic density-matrix elements $\rho_{1,1}$ (solid), $\rho_{1,0}$ (dashed), and $\rho_{1,-1}$ (dash-dotted) obtained from the Born contributions (top left), the $\N{1520}$ resonance contributions (top right), and the coherent sum of the above two assuming no extra phase factor between the $\N{1520}$ - direct $\gamma$ and $\N{1520}$ - \rhoMesonText{} terms (bottom left) and assuming a phase factor of $\rhophase = \ang{180}$ (bottom right). The lines in each set correspond to virtual photon masses $\invM=0.1$ (thinnest), 0.3, and $\gev{0.5}$ (thickest). }
    \label{fig:density_matrix}
\end{figure*}

We used our model to give numerical predictions for the density matrix elements $\rhoprod_{1,1}$, $\rhoprod_{1,0}$, and $\rhoprod_{1,-1}$. \Cref{fig:density_matrix} shows these matrix elements as a function of the scattering angle $\theta_{\gamma^*}$. We present the results obtained from the two most important contributions: the Born terms (top left) and the $s$-channel $\N{1520}$ diagram (top right). At the bottom the combined result of the above two are shown for the two cases where no extra phase factor is assumed between the direct-$\gamma$ and the \rhoMesonDash{} contributions of the $\N{1520}$ (bottom left) and a phase factor of $\rhophase = \ang{180}$ is included (bottom right). For each matrix element, three different curves are plotted assuming three different values for the invariant mass $\invM$. On the plots one can follow how the shape of the curves is influenced by the relative strength of the two dominant contributions for various invariant masses.

\section{Conclusions and Outlook}
\label{section:conclusions}
We have presented a model study of the dilepton production process in pion-nucleon collisions in the second resonance region. The applied effective-Lagrangian model is based on the earlier works of \cite{mikloswolf} and \cite{Speranza_2017}. We calculated both the contributions of baryon resonances and the non-resonant (Born) terms. For the electromagnetic interaction of hadrons, we employed a version of the vector-meson dominance (VMD) model in which hadrons couple to the electromagnetic field both directly and via an intermediate \rhoMesonText. For the interaction Lagrangians involving higher-spin resonances we used a consistent interaction scheme eliminating lower-spin degrees of freedom. We fixed the relative signs of hadron-photon and hadron-$\rhoMeson$ interaction Lagrangians based on the requirement that in the limit of universal coupling constants, the VMD model used in the present study should become equivalent to the other standard form of VMD.

We carried out numerical calculations for $\sqrt{s}=\gev{1.49}$ c.m.\ energy, which coincides with the energy of recent experiments by the HADES collaboration. The $\nucleon\rhoMeson$  coupling strengths of the baryon resonances relevant at this energy have been determined using the corresponding branching ratios that were obtained from results of the same HADES experiment on pion pair production \cite{HADESpipi}.

We presented our model predictions for the differential cross section $\mathrm{d}\sigma/\mathrm{d}\invM$ and indicated the uncertainties arising from the insufficient knowledge of the widths and branching ratios of baryon resonances. Our results show that at this energy the Born terms and the term with an $s$-channel $\N{1520}$ resonance give the most significant contributions. In the present calculation the $\N{1440}$ contribution is smaller than in the model of \rcite{Speranza_2017}, due to the very small upper limit for the $\nucleon\rhoMeson$ branching ratio of $\N{1440}$ found in \rcite{HADESpipi}.

We demonstrated how the introduction of a phenomenological phase factor for the resonance-nucleon-$\rhoMeson$ vertices influences the shape of the invariant mass spectrum. In particular, we have shown that in the absence of such a phase factor there is a constructive interference between the direct-photon and the \rhoMesonDash{} contributions of baryon resonances. The introduction of a phase factor of \ang{180} results in a destructive interference and a minimum  appears in the invariant mass spectrum around $\invM=\gev{0.4}$. A similar minimum has been seen also in \cite{mikloswolf} where a different choice was made for the signs of interaction Lagrangians involving a \rhoMesonText.

We presented predictions for elements of the spin density matrix $\rhoprod_{\lambda,\lambda^\prime}$, which represents the polarization state of the intermediate virtual photon. The density-matrix elements are given as a function of the scattering angle for various fixed values of the invariant mass. These matrix elements can in principle be determined from the angular distribution of dileptons originating from the decay of the virtual photon. However, for this the measurement of a multi-differential cross section is needed with sufficient statistics: for each fixed value of the invariant mass $\invM$ and scattering angle $\theta_{\gamma^*}$, the dependence of the differential cross section on the electron angles $\theta_e$ and $\phi_e$ will determine the density matrix elements. 

For high invariant masses ($\invM \gtrsim \gev{0.45}$), diagrams involving an intermediate \rhoMesonText{} \`a la VMD dominate over the ones containing a direct coupling to the photon, as demonstrated for the case of $\N{1520}$ contributions in \cref{fig:N1520phases}. The intermediate \rhoMesonText{} is in the same polarization state as the virtual photon it converts to. \rhoMesonsText{} decay dominantly into a pion pair, therefore the intermediate \rhoMesonText{} can be studied via the $\reactionpipi$ reaction. In particular, the polarization state and the spin density matrix of the \rhoMesonText{} might be accessible via the angular distribution of the produced pion pair. Such an analysis assumes that the \rhoMesonDash{} contribution to pion pair production can be disentangled from other contributions.

To study properties of baryon resonances in the third resonance region the HADES collaboration will continue the experimental campaign using a secondary pion beam at the c.m. energy of $\sqrt{s}=\gev{1.76}$. At this energy we expect the spin-5/2 resonances $\N{1680}$ and $\N{1675}$ to contribute to the cross section. This will most easily be visible via a different dependence of spin density matrix elements on the scattering angle $\theta_{\gamma^*}$.
Our model already includes relevant resonances for the third resonance region and can easily be extended to include missing ones. Preliminary calculations show that different choices of form factors will result in more significant differences at higher energies, therefore a careful study of their effects will be necessary.
Possible approaches to the electromagnetic form factor of hadrons relevant for higher energies include the extended vector-meson dominance model of \rcite{Krivoruchenko:2001jk}, where excited \rhoMesonDash{} states are also included, or the microscopic model of \rcite{Ramalho:2015qna,Ramalho:2016zgc,Ramalho:2020nwk}, where electromagnetic interaction of hadrons is described as a sum of valence-quark and pion-cloud contributions.

\section*{Acknowledgements}
We would like to thank Bengt Friman, Béatrice Ramstein, Piotr Salabura, Enrico Speranza, and Joachim Stroth for valuable discussions. D.N.\ was supported by the GSI F\&E.
M.B., T.G., and D.N.\ acknowledge support by the Deutsche Forschungsgemeinschaft (DFG, German Research Foundation) through the CRC-TR 211 'Strong-interaction matter under extreme conditions'– project number 315477589 – TRR 211.
MZ was supported by the Hungarian NKFIH grant no.\ 2019-2.1.6-NEMZ\_KI-2019-00001, by COST Action CA15213 "Theory of hot matter and relativistic heavy-ion collisions" (THOR), the CREMLINplus project, and the Helmholtz Alliance HA216/EMMI.

\appendix
\section*{Appendix}
\section{Lagrangians}
\label{section:lagrangians}
For the Born terms, i.e., the non-resonant contributions, we chose the same Lagrangians as \rcite{mikloswolf}. We use a pseudovector nucleon-pion coupling of the form
\begin{equation}
  \lagrangian_{\pion\nucleon\nucleon} = - \frac{f_{\pion\nucleon\nucleon}}{m_\pion}\psibar_\nucleon\gamma_5 \gamma^\li{\mu}\tauvector \psi_\nucleon \cdot \partial_\li{\mu}\pivector,
  \label{eq:L_NNpi}
\end{equation}
where $\tauvector = (\tau_1, \tau_2, \tau_3)$ are the Pauli matrices in isospin space and $f_{\pion\nucleon\nucleon}=0.97$ is a coupling constant where the value is taken from \rcite{fpiNNconstant}.
The direct coupling of the nucleon and the pion to the photon is given by the interaction terms
\begin{align}
  \lagrangian_{\photon\nucleon\nucleon} &= -e \psibar_\nucleon \left[\frac{1+\tau_3}{2}\slashed{A} - \left(\frac{1+\tau_3}{2}\kappa_\proton + \frac{1-\tau_3}{2}\kappa_\neutron\right)\frac{\sigma_\li{\mu\nu}}{4m_\nucleon}F^{\mu\nu}\right]\psi_\nucleon,\\
  \lagrangian_{\photon\pion\pion} &= -\iu e A_\li{\mu}\left(\piminus\partial^\li{\mu}\piplus - \piplus \partial^\li{\mu}\piminus\right),
\end{align}
where the anomalous magnetic moments of the proton and the neutron are $\kappa_p=1.793$ $\kappa_n=-1.913$. The interaction of the nucleon and the pion with the \rhoMesonText{} is described by Lagrangians analogous to the above two,
\begin{align}
  \lagrangian_{\rhoMeson\nucleon\nucleon} &= -\frac{\tilde{g}_\rhoMeson}{2} \psibar_\nucleon \left(\vec{\slashed{\rhoMeson}} - \kappa_\rhoMeson \frac{\sigma_\li{\mu\nu}}{4 m_\nucleon}\vec{\rhoMeson}^\li{\mu\nu}\right) \cdot \tauvector \psi_\nucleon,\\
  \lagrangian_{\rhoMeson\pion\pion} &= - \tilde{g}_\rhoMeson \left[(\partial^\mu\pivector)\times\pivector\right] \cdot \vec{\rhoMeson}_\li{\mu},
\end{align}
where a value of $\tilde{g}_\rhoMeson = 5.96$ is obtained from the width of the $\rhoMeson\to\pi\pi$ decay. For the tensor coupling, $\kappa_\rhoMeson$, values between 1.99 and 2.65 have been obtained in \rcite{Feuster:1997pq} based on various PWA solutions. Here we use $\kappa_\rhoMeson = 2.3$.
The presence of the derivative of the pion field in the nucleon-pion interaction, Eq.~(\ref{eq:L_NNpi}) forces us to introduce four-point interactions of the form
\begin{align}
  \lagrangian_{\photon\pion\nucleon\nucleon} &= -\frac{\iu e f_{\pion\nucleon\nucleon}}{m_\pion}\psibar_\nucleon \gamma_5 \gamma^\li{\mu} \tauvector \psi \cdot A_\li{\mu} Q \pivector,\\
  \lagrangian_{\rhoMeson\pion\nucleon\nucleon} &= - \frac{\tilde{g}_\rhoMeson f_{\pion\nucleon\nucleon}}{m_\pion}\psibar_\nucleon \gamma_5 \gamma^\li{\mu} \tauvector \psi \cdot \left(\vec{\rhoMeson}_\li{\mu} \times \pivector\right)
\end{align}
in order to maintain gauge invariance.

In the present model we describe the baryon resonance transition vertices by the same Lagrangians as in \rcite{Speranza_2017}, but we also include direct coupling terms to the photon, using forms analogous to the corresponding interaction Lagrangians with the \rhoMesonText{}. For higher-spin resonances we employ the consistent interaction scheme of \rcite{Vrancx_2011}. The interaction Lagrangians involving spin-1/2 and spin-3/2 resonances are given by
\begin{align}
    \lagrangian_{\piNR_{\oneHalf}} &= -\frac{2\couplingConstantPiNR}{m_\pion}\psibar_\resonance\varGammaFive\gamma^\li{\mu}\mathcal{\xvec{T}}\psi_\nucleon\cdot\partial_\li{\mu}\pivector+\hermitianConjugate\\
    \lagrangian_{\piNR_{\threeHalves}} &= \im\frac{2\couplingConstantPiNR}{m_\pion^2}\Psibar_\resonance^\li{\mu}\varGammaFive\mathcal{\xvec{T}}\psi_\nucleon\cdot\partial_\li{\mu}\pivector+\hermitianConjugate\\
    \lagrangian_{\rhoNR_{\oneHalf}}&=\frac{\couplingConstantRhoNR}{m_\rhoMeson}\psibar_{\resonance}\mathcal{\xvec{T}}\diracSigma^\li{\mu\nu}\varGammaFiveTilde\psi_\nucleon\cdot\vec{\rhoMeson}_\li{\mu\nu}+\hermitianConjugate\\
    \lagrangian_{\rhoNR_{\threeHalves}} &=-\im\frac{\couplingConstantRhoNR}{2m_\rhoMeson^2}\Psibar_\resonance^\li{\mu}\mathcal{\xvec{T}}\gamma^\li{\nu}\varGammaFiveTilde\psi_\nucleon\cdot\rhovector_\li{\mu\nu}+\hermitianConjugate\\
     \lagrangian_{\photonNR_{\oneHalf}}&=-\frac{\couplingConstantGammaNR}{2m_\rhoMeson}\psibar_{\resonance}\diracSigma^\li{\mu\nu}\varGammaFiveTilde\psi_\nucleon F_\li{\mu\nu}+\hermitianConjugate\\
    \lagrangian_{\photonNR_{\threeHalves}} &=\im\frac{\couplingConstantGammaNR}{4m_\rhoMeson^2}\Psibar_\resonance^\li{\mu}\gamma^\li{\nu}\varGammaFiveTilde\psi_\nucleon F_\li{\mu\nu}+\hermitianConjugate
\end{align}
where $\resonance_{J}$ corresponds to a baryon resonance with total spin $J$.
$\varGammaFive=\gamma^5$ if $J^P \in \left\{1/2^+,3/2^-\right\}$ and $\varGammaFive=1$ otherwise,  and $\varGammaFiveTilde = \gamma^5\varGammaFive$. $\mathcal{\xvec{T}}=\tauvector/2$ if the baryon resonance 
has a total isospin equal to $1/2$ ($\resonance = \nucleon^*$), while $\mathcal{\xvec{T}}=\Thetavector$
in case of 
an isospin-3/2 resonance ($\resonance = \Delta^*$).
Here $\Thetavector = (\Theta_1, \Theta_2, \Theta_3)$ are the isospin  1/2 to 3/2 transition matrices.

In line with the consistent scheme of \rcite{Vrancx_2011}, spin-3/2 baryons are represented by the field
\begin{equation}
    \Psi_\mu = i\gamma^\nu(\partial_\mu\psi_\nu - \partial_\nu \psi_\mu),
\end{equation}
where $\psi_\mu$ is the traditional Rarita-Schwinger field.

The propagators for the spin-1/2 particles are given by
\begin{align}
 S_{\oneHalf} &=     
 \frac{i}{p^2_R-m^2_R+i\sqrt{p^2_R}\Gamma_R}(\slashed{p}_R+m_R)
\end{align}
and those for the spin-3/2 particles by
\begin{align}
S^{\mu\nu}_{\threeHalves} &= \frac{i}{p^2_R-m^2_R + i \sqrt{p_R^2}\Gamma_R(p_R^2)}P^{\mu\nu}_{\threeHalves}(p_R,m_R),
\end{align}
with the projector
\begin{align}
    P^{\mu\nu}_{\threeHalves} &= -(\slashed{p}_R + m_R)\left(\mt^{\mu\nu} - \frac{\gamma^\mu\gamma^\nu}{3}\right).
\end{align}
Terms in the projectors that are proportional to $p^\mu_R$ have been neglected since they cancel out when contracted with the vertices. The general expression can be found in \rcite{mikloswolf}. Furthermore we use the same width parametrization for baryon resonance as in \rcite{mikloswolf}. 


\section{Coupling Constants}
\label{sec:couplingConstants}
We took the Breit-Wigner mass $m$ and total decay width $\Gamma$ of baryon resonances from the Review of Particle Physics \cite{pdg} by the Particle Data Group (PDG). The PDG provides an upper ($u$) and lower ($l$) value for the total widths of resonances and their branching ratios ($BR$) to the $\nucleon\pion$ and $\nucleon\photon$ final state. We used $BR=\frac{u+l}{2}\pm\frac{u-l}{2}$ for the mean value and the uncertainty of the branching ratios and a similar expression for the total widths. Using these we obtain the partial widths and their uncertainties. The values and uncertainties of the corresponding coupling constants are determined by requiring that our model's prediction for the partial widths reproduces the PDG values and uncertainties.
 
Using a partial-wave analysis of the reaction $\piminus\proton \to \nucleon\pion\pion$, the HADES experiment reported for the branching ratios of $\N{1520}$ and $\N{1535}$ to the $\nucleon\rho$ final state the values $(12.2\pm 1.9)\%$ and $(3.2\pm 0.7)\%$, respectively \cite{HADESpipi}, while for the $\N{1440}$ an upper limit of $\SI{0.2}{\percent}$ was obtained, which we interpret as $(0.1\pm0.1)\%$. The coupling strength of the resonance to the $\nucleon\rhoMeson$ channel can be determined based on the decay mode $\resonance\to \nucleon\rhoMeson \to \nucleon\pi\pi$. From the effective-Lagrangian model one can obtain the prediction for this decay width via integration over the \rhoMesonDash{} spectral function according to
\begin{multline}
    \Gamma_{\resonance\to \nucleon\pi\pi(\rhoMeson)} = \Gamma_{\resonance\to \nucleon\rhoMeson} = \\ \frac{1}{4\pi^2}\frac{|\mathbf{p}_{\nucleon}|}{m_\resonance^2}
    \int \mathrm{d}\invM \frac{1}{n_{\text{pol}}} \sum|\mathcal{M}_{\resonance\to \nucleon\rhoMeson}|^2
    \frac{\invM^2 \Gamma_{\rho}(\invM)}{(\invM^2-m_{\rhoMeson}^2)^2 + \invM^2{\Gamma_{\rho}(\invM)}^2},
\end{multline}
where we have taken into account that the \rhoMesonText{} decays to a pion pair with $\approx 100\%$ branching ratio.
The $\rhoNR$ coupling constants and their uncertainties are determined by requiring that the model reproduces the $\nucleon\rhoMeson$ partial widths obtained by combining the total width given by PDG and experimental $\nucleon\rhoMeson$ branching ratio by HADES. The coupling constants of baryon resonances are summarized in \Cref{tbl:resonanceOverview}.

\printbibliography
\end{document}